\documentclass[fleqn,10pt]{wlscirep}
\usepackage[utf8]{inputenc}
\usepackage[T1]{fontenc}
\title{Improving Magnetic Resonance Imaging with 
Smart and Thin Metasurfaces}

\author[1]{Endri Stoja}
\author[2]{Simon Konstandin}
\author[2]{Dennis Philipp*}
\author[2]{Robin N.\ Wilke}
\author[1]{Diego Betancourt}
\author[1]{Thomas Bertuch}
\author[2,3]{J\"urgen Jenne}
\author[2,3]{Reiner Umathum}
\author[2,4]{Matthias G\"unther}

\affil[1]{Fraunhofer FHR, Fraunhoferstraße 20, 53343 Wachtberg, Germany}
\affil[2]{Fraunhofer MEVIS, Max-von-Laue-Str. 2, 28359 Bremen, Germany}
\affil[3]{Division of Medical Physics in Radiology, German Cancer Research Center DKFZ,
Im Neuenheimer Feld 280, 69120 Heidelberg, Germany}
\affil[4]{MR-Imaging and Spectroscopy, Faculty 01, University of Bremen, Otto-Hahn-Allee 1, 28359 Bremen, Germany}

\affil[*]{dennis.philipp@mevis.fraunhofer.de}

\keywords{Magnetic Resonance Imaging, Metamaterials, Metasurfaces, signal-to-noise ratio, field enhancement}

\begin{abstract}
Over almost five decades of development and improvement, Magnetic Resonance Imaging (MRI) has become a rich and powerful, non-invasive technique in medical imaging, yet not reaching its physical limits. 
Technical and physiological restrictions constrain physically feasible developments. 
A common solution to improve imaging speed and resolution is to use higher field strengths, which also has subtle and potentially harmful implications. 
However, patient safety is to be considered utterly important at all stages of research and clinical routine. 
Here we show that dynamic metamaterials are a promising solution to expand the potential of MRI and to overcome some limitations. 
A thin, smart, non-linear metamaterial is presented that enhances the imaging performance and increases the signal-to-noise ratio in 3T MRI significantly (up to eightfold), whilst the transmit field is not affected due to self-detuning and, thus, patient safety is also assured. 
This self-detuning works without introducing any additional overhead related to MRI-compatible electronic control components or active (de-)tuning mechanisms. 
The design paradigm, simulation results, on-bench characterization, and MRI experiments using homogeneous and structural phantoms are described. 
The suggested single-layer metasurface paves the way for conformal and patient-specific manufacturing, which was not possible before due to typically bulky and rigid metamaterial structures. 
\end{abstract}

\begin{document}

\flushbottom
\maketitle

\thispagestyle{empty}

\section*{Introduction and Motivation}

Magnetic Resonance Imaging (MRI)\cite{Lauterbur:1973,Nishimura:2010} is the most versatile and powerful imaging modality available for clinical use nowadays. 
For almost five decades, the technology was improved, extended, and continues to evolve, still not reaching the physical limits. 
Technical and physiological limitations are hampering the advancement and constrain physically feasible developments, making it increasingly challenging to innovate.
 
While in clinical applications static magnetic field strengths of 1.5 and 3 Tesla are most common, research scanners have been developed with field strengths of $7\,$T, $9.4\,$T, and even higher to benefit from the increase in signal-to-noise ratio (SNR).\cite{Cao:2015,Ocali:1998}
However, working at higher field strengths becomes progressively difficult due to the direct proportionality to the Larmor frequency, which defines the nuclear resonance frequency of (hydrogen) atoms. 
A key limiting factor is the specific absorption rate (SAR) of the deposited radio frequency (RF) power that increases almost quadratically with frequency, setting practical limitations due to tissue heating.\cite{Schick:2005}
Besides general challenges related to strong magnetic fields, electromagnetic wave phenomena become more relevant due to the shorter wavelength of the RF field.
Huge efforts have to be taken to tackle issues arising from the fact that the RF wavelength is on the order of the imaged object’s dimensions, eventually creating interference patterns and standing waves.\cite{Caserta:2004} 
As a consequence, MRI is sensitive to motion artifacts, which should be circumvented.
At higher field strengths, the longitudinal relaxation time $T_1$ of tissue increases, leading to an intrinsic limitation on the achievable imaging speed for some applications.\cite{Zhang:2013} 

A main technical (and physiological) obstacle in MRI is also the use of gradient magnetic fields for localization, which substantially limit the technically possible imaging speed.\cite{Ham:1997,Reilly:1998}
Despite recent advances in fast imaging approaches\cite{Uecker:2010} such as parallel imaging\cite{Griswold:2002,Pruessmann:1999} and compressed sensing,\cite{Lustig:2007} image acquisition can be considered comparably slow. 
To improve the imaging performance, the patient-specific design of MRI equipment certainly is a possible approach\cite{Corea:2016} but usually related to high costs as well as complicated design and manufacturing when it comes to, e.g., tailored receive coil arrays.

The best solution strategy to enhance the MRI performance and proceed to the next level of medical imaging should offer increased imaging efficiency (a metric of SNR, contrast, and speed) over large volumes of interest without the immediate need of higher background field strengths. 
So-called metamaterials (MTMs) are an up-and-coming solution in this respect. 
Here we show that a smart metasurface - a dynamic two-dimensional MTM - yields a significant SNR improvement for $3\,$T MRI without affecting the transmit field. 
Electromagnetic MTMs, as outlined by the seminal works of Vesselago, Pendry, Schurig, Leonhardt, and others\cite{Veselago:1968,Pendry:2006,Schurig:2006,Leonhardt:2006,Leonhardt:2006b} are artificially constructed structures, consisting of a usually periodic arrangement of dielectric or conducting unit elements (small $L$-$C$ circuits). 
Prominent application examples include MTM cloaks, perfect lenses, and ultimately MRI applications\cite{Solymar:2009}. 
MTMs offer to manipulate the amplitude (focusing effects) as well as the phase of incident radiation \cite{Yuan:2020_rev1, Zhang:2020_rev2} and also polarization-independent modulation can be achieved \cite{Yuan:2019_rev3}.
The unit cells can be considered as single meta-atoms on a sub-wavelength scale. 
Therefore, an incident electromagnetic field is subject to a macroscopic influence induced by the interactions of all meta-atoms. 
Hence, w.r.t.\ RF field interaction the MTM can be viewed as a homogeneous material slab, which is effectively described by (anisotropic and dispersive) permeability and permittivity. 
In contrast to naturally occurring materials, MTMs can be designed to have arbitrary positive and negative values for both parameters. 
This leads to, e.g., field enhancement or focusing, phase changes, and tailored reflection and transmission properties. 
Usually, metasurfaces yield desired effects only for a specific target frequency with a very narrow bandwidth. 
This property makes them ideal tools for MRI, which similarly builds on narrow bandwidth RF radiation. 
It was shown previously that MTMs can be used to significantly improve the SNR without the need of stronger background fields.\cite{Schmidt:2017,Duan:2019,Freire:2010,Kretov:2018,Kretov:2019,Shchelokova:2018,Shchelokova:2018b,Slobozhanyuk:2016}
However, passive structures as those presented in the available literature will of course influence both, the transmit (Tx) as well as the receive (Rx) field. 
The influence during Tx corresponds to a local increase of power transmitted, posing a potential threat of tissue heating and high SAR values. 
Moreover, obviously the Tx-field will be disturbed leading to a range of unwanted effects and subtle consequences.
Thus, it is utterly important to carefully design, test, and use MTMs in MRI applications as patient safety must be assured at all times in research as well as clinical routine. 
To do so, dynamical MTMs can be developed, which are (de-)tuned depending on the imaging phase, i.e., discrete or continuous resonance states are to be introduced. 
This is indeed not trivial due to the MRI scanner's strong magnetic field environment, which limits electromagnetic signal communication close to the machine and causes some known control mechanisms for MTMs not to work. 
So far, only very few tunable MRI MTMs have been presented in general, and there are even less proposals that achieve the tuning depending on the imaging phase without the need of active intervention.\cite{Zhao:2019,Saha:2020} 
Non-linear MTMs\cite{Zhao:2019} are introduced as a promising solution since they can be sensitive to the imaging phase but also MRI-compatible additional battery-powered electronic sensing circuits\cite{Saha:2020} can be used to take care of the state-switching. 
However, in all known cases the suggested MTMs are bulky structures (such as helix-shaped configurations or 3d arrangements of conducting bars), putting patient-specific design, patient comfort, and flexible or conformal manufacturing benefits beyond reach. 
The thickness of several centimeters also limits possible imaging applications.
To overcome these limitations, we developed smart and ultra-thin metasurface enhancement plates (EPs) for MRI. 
Here, ``smart'' refers to automatic self-detuning in the Tx phase of the imaging.
The functionality was designed and tested via simulations and verified in on-bench experiments and MRI scans at $3\,$T. 
Depending on the imaging sequence parameters, the SNR is increased up to eightfold (by a conservative measure) in slices close to the metasurface. 
A generalization to different field strengths is easily possible, such that EPs can be constructed for, e.g., $1.5\,$T or $7\,$T scanners.

Note that the MTMs (in general and those presented here in particular) do not act as MRI coils.
Rather their effect is to locally redistribute (focus) the Rx field (or Tx field in other applications) such that in combination with the scanner's Tx and Rx coils the SNR can be enhanced, leading to increased imaging efficiency.
The smart metasurfaces here are designed to be used in combination with the MRI scanner's integrated body coil.
Thus, they allow to build a kind of effective, local, universal (referred to different imaging applications with the same metasurface), and wireless Rx coil, which is more cost-effective and simpler w.r.t.\ dedicated Rx coils.
MTMs which work in combination with other (local) Rx coils can also be designed but care needs to be taken concerning the interaction of close resonant structures.
However, as resonant, conducting structures are involved in the construction, the design of MTMs for use in MRI applications may adopt and modify conventional Rx/Tx coil methodologies and well-known geometries.

\section*{Results}
\subsection*{Metasurface design paradigm and (de-)tuning mechanism}
In the following, we present the final results of our iterative metamaterial design loop for smart metasurfaces.
To avoid any additional control circuitry components and batteries, we follow a  design paradigm for non-linear MTMs\cite{Zhao:2019}, such that each of our smart metasurface EPs consists of a non-linear system of two inductively coupled and resonant substructures. 
One of these is a linear 2d MTM (linear metasurface) and the other one is a non-linear, single-loop tuning resonator; see Figs.\ \ref{fig:1}d, \ref{fig:2}a and the supplementary material for a visual impression. 
In such an arrangement, resonant hybrid modes\cite{Jouvaud:2016} are determined by the properties of each subsystem and the coupling mechanism. 
The fundamental mode is of primary interest for MRI applications due to the largest region of interest (ROI) that can be covered. 
Our design has the advantage that instead of manipulating every unit cell of the linear MTM separately, only the tuning resonator needs to be controlled to vary the resonance of the full EP. 
Details on a theoretical description of a similar arrangement via the coupled mode theory have been published before\cite{Zhao:2019} and we outline the most important steps in the methods section; see also the supplementary material and figures therein.
Details on the design parameters for the EPs are outlined in the methods section as well. 

Following the design methodology, each of our EPs is conceptually composed of i) a linear metasurface resonator consisting of capacitively-loaded and coupled flat-wire unit cells, and ii) an outer single-loop tuning resonator, which is milled on the same dielectric substrate and tightly covers the inner structure. 
The substrate is less than $0.6\,$mm thick with only $17\,\mu$m of copper cladding, see the methods section for details. 
The tuning resonator is loaded with semi-conductor elements that are sensitive to the incident power level via the induced current in the loop.
In the Tx phase of the MRI sequence, substantially higher power is incident as compared to the Rx phase, in which the signal is caused by the relaxation of the excited magnetization. 
Hence, the resonance behavior of the tuning resonator is sensitive to the imaging phase and, thus, also the hybrid modes of the full EP are influenced by the incident power. 
The result is a smart, non-linear metasurface of which the resonance properties change between Tx and Rx without any additional need for user influence, control circuitry, batteries, or active (de-)tuning. 

The resonance at low incident power is made to coincide with the MRI scanner’s Larmor frequency (123.5 MHz here, see the methods section) by the design of the metasurface such that the EP is active and resonant in the Rx phase while it is sufficiently detuned (almost invisible) in Tx. 
For the first version (EP1), varactor diodes are soldered in series into the tuning resonator, while in the second case (EP2) an MRI compatible limiter diode is used, see Fig.\ \ref{fig:3}e. 
To provide the possibility of manual fine-tuning of the resonance frequency at low incident power levels, a small trimmer capacitor was soldered into the tuning resonator in either case. 
It allows to tune and characterize the smart metasurface on-bench in presence of a phantom before using it in the MRI scanner. 
For the inner metasurface, the wire-resonator unit cells are loaded with capacitors, which are implemented as simple rectangular patches on a ground plane (on the back layer, see the figures in the supplementary material). 
In this way, the intensive use of lumped elements can be avoided to increase the Q-factor. 
In a next step, the manufacturing on flexible substrates for patient-specific and/or conformal design for many different applications becomes possible due to the compactness of the metasurfaces, the structural design that avoids lumped elements, and laser milling on thin sub-millimeter substrates. 
This is a big advantage over existing (bulky) solutions that use helix-structures or 3d arrangements of wire-shaped conducting bars with a thickness of several centimeters in total.
Note that our metasurface resonator itself resembles properties of an open low-pass birdcage coil with the leg capacitors being a series of the two structural parallel-plate ones at the wire ends. 
The metasurface design was successively iterated based on simulations and measurement results, which are presented in the next sections.

\subsection*{Simulation, and on-bench characterization results}
For the $B_1$ field enhancement in MRI, we are interested in the lowest-order resonant hybrid mode of the EP. 
Eigenmode simulations were conducted to study the resonant modes of the substructures at first separately and then as a coupled system. 
The lowest-order coupled modes involving the fundamental mode of the inner, linear metasurface are two: one in which the fields generated by the substructures are in phase (mode of interest), whereas in the other one they are out of phase, see Fig.\ \ref{fig:1}a. 
In the next step, a plane wave excitation was used for full-wave simulations of the entire structure in presence of a cylindrical phantom to assess the characteristics and to feedback the design loop. 

After successive design cycles, prototypes are manufactured for on-bench measurements of i) the resonance frequency as a function of the incident power, and ii) the quality factor to assure the desired functionality. 
A typical setup composed of shielded sniffer coils connected to a vector network analyzer was used, see Fig.\ \ref{fig:1}d and the supplementary material. 
For $S11$ measurements at different incident power levels, the sniffer coil is positioned some distance above the center of the EP. 
The self-detuning of the smart metasurface is clearly observable at increasing incident power, see Fig.\ \ref{fig:1}e. 
The same setup is used to fine-tune the EP at the target frequency at low incident power by variation of the trimmable capacitor. 
For the quality factor we find ${Q = 110}$ for EP1 and ${Q = 103}$ for EP2. 
In Fig.\ \ref{fig:1}f, the detuning of the EP as a function of the incident power is shown in terms of S21 measurements with the sniffer coils positioned at a distance of $300\,$mm from each-other and the EP in the center. 
This detuning can be observed as a shift towards lower frequencies of the maximum of $S21$ magnitude or as the drop of the $S21$ magnitude at a fixed target frequency. 
In addition, a custom-made 3-axis scanning stage was used to further characterize the spatial dependence of the $S$-parameter response as a function of incident power as shown in Figs.\ \ref{fig:1}b and c; see also the supplementary material for the setup. 
In the non-detuning regime (below about $-15\,$dBm input power), we observe an almost exponential decay of the field enhancement in the direction orthogonal to the EP. 
Depending on the input power, enhancement factors of about 10 can be observed close to the surface in free space, see Fig.\ \ref{fig:1}.

\subsection*{Magnetic resonance images and validation}
MRI measurements prove the expected functionality of the smart metasurfaces. 
A gradient echo sequence is used with the following parameters: ${TR = 100\,}$ms, ${TE = 5\,}$ms, field-of-view = $\{ 64\times 64, 128 \times 128 \}\,$mm$^2$, matrix size = ${128\times128}$, bandwidth = $250\,$Hz/Px, eight slices with thickness of $5\,$mm and distance of $15\,$mm. 
The imaging is done with a homogeneous cylindrical phantom (with both, EP1 and EP2) and with a Kiwi fruit for structural images (with EP1). 
For high SNR images of the Kiwi, also measurements with increased ${TR = 1\,}$s are performed. 
The scanner’s body coil is used for Tx throughout all experiments and also for Rx in combination with the smart metasurfaces EP1 and EP2. 
For comparison, imaging is also done with a local single loop coil (SLC) for Rx while the EP is removed. 
Besides the integrated body coil of the MRI scanner, a SLC might be considered a standard to which a new MTM device should be compared in a first step.

All measurements are performed with a single sequence implemented into the vendor-independent gammaSTAR framework (Fraunhofer MEVIS)\cite{Cordes:2019, GammaStar} with dedicated pauses between measurements to avoid frequency adjustments between successive experiments with/without the metasurfaces, see also the methods section. 
To investigate the de-tuning behavior and to separate effects on the Tx and Rx fields, 10 measurements with equidistant nominal flip angles between 0 and 90 degrees are recorded to determine the Ernst angle, i.e., the flip angle leading to a maximal SNR. 
This approach allows to genuinely characterize MTMs in MRI applications and to clearly differentiate between Tx and Rx modifications of the $B_1$ field, see the methods section, but this method is often missing in existing literature. 
Reference measurements are performed without any metasurface in either case using only the body coil. 
The geometrical setup containing the phantom and the smart metasurface is shown in Fig.\ \ref{fig:2}. 

\begin{figure}
  \centering
  \includegraphics[width=\linewidth]{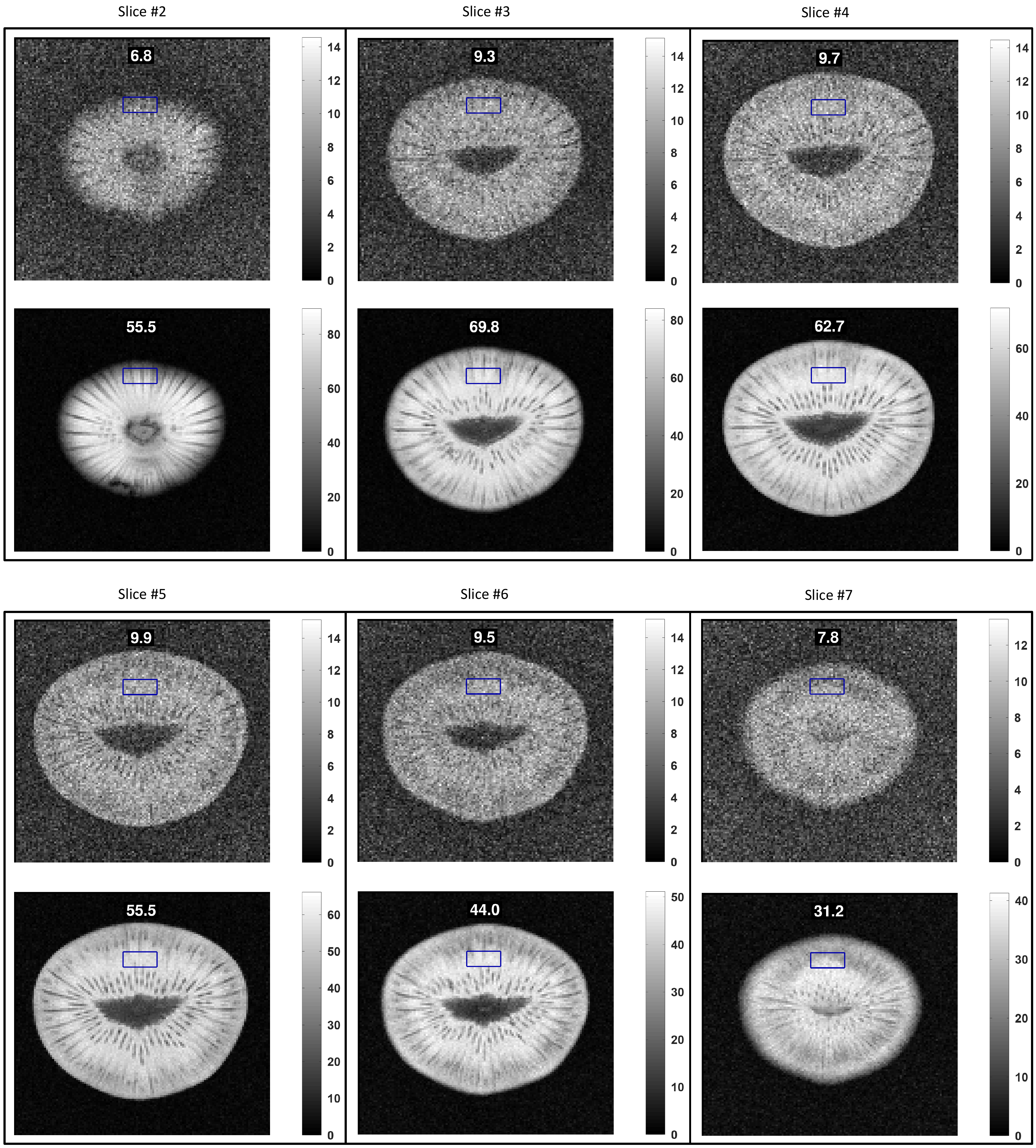}
  \caption{Additional MRI results for structural images with the Kiwi fruit and
${TR = 1\,}$s. The plots show the comparison of results for the body coil (top) and the body coil + smart metasurface (bottom) in each panel. The flip angle is the approx.\ Ernst angle in this configuration. The respective SNR in the ROI is indicated in the subfigures.
}
  \label{fig:6}
\end{figure}

The noise for all SNR evaluations is taken from the respective $0\,$deg flip angle measurement, which is a better and more conservative measure than taking the noise from apparently “signal-free” areas in the actual images. 
The results in Fig.\ \ref{fig:3} and Fig.\ref{fig:4} prove that the detuning mechanism works well for both EPs. 
The Ernst angle is the same for measurements with and without the smart metasurfaces, see Fig.\ \ref{fig:4}d and the supplementary material.
Hence, in the Tx phase of the imaging sequence, the EPs remain silent and do not affect the $B_1$ field. 
In the Rx phase, however, they lead to a sixfold increase in SNR for homogeneous phantom measurements with ${TR = 100\,}$ms and to an almost eightfold SNR increase for structural measurements with ${TR = 1\,}$s, see Figs.\ \ref{fig:3}, \ref{fig:4}, \ref{fig:5}, and \ref{fig:6}.
The homogeneous phantom results in Fig.\ \ref{fig:3} show that both EPs work approximately equally well. 
From Fig.\ \ref{fig:3}f, one can see that after a certain penetration depth, roughly the diameter of the SLC, the metasurfaces yield larger SNR improvements and outperform the local coil. 
Moreover, the profile for the SLC decreases almost exponentially, whereas the enhancement factor due to the smart metasurfaces seems to decrease only linearly. 
This behavior is confirmed in simulations and offers yet another advantage compared to local coils and existing MTM solutions.

\section*{Discussion}
The metasurfaces introduce local focussing effects that may be compared in some sense to effects of a lens. 
However, note that our metasurface design differs from what is usually called “metamaterial lens”. 
The smart metasurface's functionality might be best described in terms of a local resonant structure (composed of an array of wire-shaped unit cells) exited by the Rx field, which inductively couples to the body coil resonator of the MRI-scanner. In this case, we have near-field coupling in contrast to far-field coupling by, e.g., a lens.
Then, the incident field induces a current in the single unit cells (Faraday’s law of induction) and, in turn, the currents flowing in the unit cells induce magnetic fields superimposed on the incident one. 
The respective currents depend on the material, the mutual coupling and the geometry - these parts we can influence by the metasurface design. 
The full enhancement plate has multiple resonant modes, of which the ground mode is of major interest here. 
It also depends, as shown by the theoretical description, on the coupling to the non-linear outer tuning loop.

Compared to the existing solutions for detunable MRI MTMs\cite{Zhao:2019,Saha:2020}, our smart metasurfaces exhibit a larger penetration depth and slower decrease of the SNR enhancement factor while maintaining a similar or even better performance. 
It is a subtle problem though to make a quantitative comparison as it depends on, e.g., the size of the respective MTMs, the imaged object, and the MRI sequence. 
From looking at SNR enhancement only, other authors claim an improvement factor of about 15\cite{Zhao:2019}. 
However, for this result a small $2\times1$ configuration with (just) two helix-shaped unit cells is used. 
Of course, this smaller configuration can yield higher local SNR effects very close to the metamaterial but these are rapidly dropping as distance increases. Thus, the enhancement is at the expense of a larger field-of-view.
As the field-of-view in MRI (for most applications) is larger than this $2\times1$ configuration, a larger MTM structure - such as the one presented here - is needed accordingly. 
In the supplementary material of the paper, the authors show results for a $4\times4$ configuration of their MTM, which is pretty bulky in size; see Figure S3 in their supplementary material. 
For this configuration, which compares roughly to our thin metasurface in the planar dimensions, a SNR enhancement of about 6-7 is reported but also a different repetition time of TR = $3\,$s is used for the MRI sequence. 
Note that our smart metasurfaces yield a similar or even better enhancement with TR = $\{100,1000\}\,$ms.

Our results clearly show that the SNR in MRI can be substantially improved with smart yet thin MTMs. 
Safety concerns are taken care of by the automatic Tx detuning and the sub-millimeter structures set the foundation for the possibility of flexible and conformal manufacturing. 
In some cases, the metasurfaces in combination with the scanner’s body coil even outperform a local SLC. Future work devoted to the improvement and construction of even better and advanced MTMs will also address a detailed comparison to local Rx coils and the potential use of MTMs as their (partial) replacement in imaging applications.
If the same image quality can be achieved without local coils, which are wired to the scanner’s interface, metasurfaces have the potential to improve patient comfort and function effectively as ``universal wireless receive coils''.

\section*{Methods}
\subsection*{Simulation, metasurface design, and theoretical modeling}
All numerical electromagnetic simulations were performed using CST Microwave Studio (Dassault Syst{\`e}mes, France) with a frequency domain solver. 
Two different types of simulations, eigenmode analysis and full plane wave excitations, are employed. 
The overall design requirement is to have the full smart metasurface resonant in the Rx phase of the imaging process at the MRI scanner’s Larmor frequency, which is 123.5 MHz in our case ($3\,$T Magnetom Skyra, Siemens Healthineers, Germany). 
To achieve this goal, the wire-resonator metasurface can be varied in its overall dimension and the length of substructures as well as the number of wire resonators can be adjusted. 
Changing the size of single wire resonators, the MTM unit cells, allows to shift the resonance frequency. 
Furthermore, the variation of the end parts, the parallel plate capacitors, allows to vary the resonance and coupling of the unit cells. 
Since the full structure gives rise to hybrid eigenmodes, also the coupling and the properties of the outer, non-linear tuning resonator change the overall resonance frequency. 
Hence, parameter studies and optimization were performed to find the best possible combination. 
For the two manufactured EPs, we use ${N = 14}$ wire-resonator unit cells, which are $10\,$mm wide each, with printed wires of $1\,$mm width.
The capacitive end patches have the same size for the two EPs, ${9\times3\,}$mm$^2$, see Fig.\ \ref{fig:1} and the supplementary material. The length of an individual unit cell is $180\,$mm.
Each EP consists of two subsystems: the inner, linear metasurface and the outer, non-linear tuning resonator. 
These two subsystems were designed to individually resonate at slightly higher frequencies, respectively, so that when inductively coupled together they would resonate at the target frequency of the MRI system. 

For the simulations, semiconductor components (varactor diodes for EP1 and the limiter diode for EP2), are modeled as fixed-value capacitors (lumped elements) with different discrete values for Tx and Rx, respectively. 
Full wave simulations are performed with open boundary conditions in all directions and plane wave excitations with circular polarization.
The propagation direction of the plane wave is in the plane of the metasurface along the symmetry axis orthogonal to the wire resonators. 
A homogeneous phantom of which the properties such as density, permittivity, and conductivity resemble average values for tissue ($\epsilon_r = 78$, $\sigma = 0.7 \,$S/m) was included to simulate the loading. 
Evaluations of the numerical results show that the $H$-field at the target frequency inside the phantom is stronger in the resonant case (corresponding to the Rx phase of MRI) and almost unaffected in the detuned case (corresponding to the Tx phase of the MRI sequence). 
Eigenmode simulations indicate that two different hybrid modes exist that involve the fundamental mode of the linear metasurface, one of which has the two subsystems in phase, see Fig.\ \ref{fig:1}. 
For the other mode, the two are exactly out of phase but the homogeneity is improved in the central part. 
For MRI, the first mode is of primary interest due to the larger region of interest covered.

An approximate theoretical description of the smart metasurfaces can be obtained using the coupled mode theory as outlined in the available literature.\cite{Joannopoulos:book, Zhao:2019}
For the linear part of the system, the inner metasurface structure, the mode amplitude $y_1$ depends on an excitation $x$ according to
\begin{align}
	\dot{y}_1 = a_1(\omega_1) \, y_1 + b_1 \, x \, , \quad a_1(\omega_1) = i \omega_1 - 1/\tau_r - 1/\tau_i \, , \quad b_1 = \sqrt{2/\tau_r} \, .
\end{align}
Here, $\omega_1$ is the resonance frequency, $\tau_r$ is the decay time due to radiation losses, and $\tau_i$ is the decay time associated with intrinsic losses. The overdot denotes the time derivative.
A similar equation can be constructed for the tuning resonator,
\begin{align}
	\dot{y}_2 = a_2(\omega_2) \, y_2 + b_2 \, x \, ,
\end{align}
but now the resonance frequency $\omega_2$ depends on the resonator's amplitude as well, $\omega_2 = \omega_2(y_2)$.
We may assume a linear relation such that $\omega_2 = \Omega_2 + \kappa |y_2|$ with a constant $\kappa$ that depends on the properties of the chosen non-linear element and a small amplitude limit $\Omega_2$.
Finally, the full metasurface is described by the coupled, non-linear system of differential equations in the time domain,
\begin{subequations}
\begin{align}
	\dot{y}_1 &= \big( i \omega_1 - 1/\tau_{r,1} - 1/\tau_{i,1} \big) y_1 + \sqrt{2/\tau_{r,1}} \, x + i k y_2 \, , \\
	\dot{y}_2 &= \big( i \Omega_2 + i\kappa |y_2| - 1/\tau_{r,2} - 1/\tau_{i,2} \big) y_2 + \sqrt{2/\tau_{r,2}} \, x + i k y_1 \, ,
\end{align}
\end{subequations}
which includes the coupling strength $k$. 
In the supplementary material details on the solution are outlined and it is shown how to qualitatively compare the reflection of the system to the S11-measurements we performed with an untuned sniffer coil, see Fig.\ \ref{fig:1}e.

\subsection*{Manufacturing, on-bench measurements and MRI scans}
The metasurfaces are manufactured at the Fraunhofer Institute for High Frequency Physics and Radar Techniques (FHR) by laser milling on Rogers 4003c substrate with a $17\, \mu$m copper layer on both sides. 
The tuning resonator loop used for automatic detuning during the Tx phase is milled on the same layer as the capacitively-loaded wires. 
The substrate’s thickness is $0.508\,$mm. For EP1, the varactor diode-loaded tuning loop has three readily available Skyworks SMV2020 varactor diodes in series to have it resonate at a suitable frequency while keeping the geometrical dimensions adjusted to the inner substructure. 
The safe use of these varactor diodes in a 3 T experiment was previously reported.\cite{Zhao:2019} 
However, varactor diodes connected in series give rise to (unwanted) non-linear effects.\cite{Powell:2007,Wang:2008}
For EP2, the limiter diode-loaded loop is of the same size as the varactor-loaded loop of EP1 but uses the UMX9989AP diode (Microsemi, USA). 
For low incident power, this limiter diode has a typical capacitance of about $4\,$pF. 
For high incident power, the limiter diode conducts and, thus, on/off states are realized as extreme cases depending on the incident power level, i.e., depending on the MRI imaging phase. 
In the two extreme states, the resonance frequency of the metasurface is different, with the low incident power state’s resonance frequency matched to the MRI scanner’s Larmor frequency. 
On the back layer of the EPs, the ground stripes for the capacitively coupled wire resonators of the inner metasurface are milled, see the supplementary material for an overview of front and back layers of the two manufactured prototypes.

\begin{figure}
  \centering
  \includegraphics[width=\linewidth]{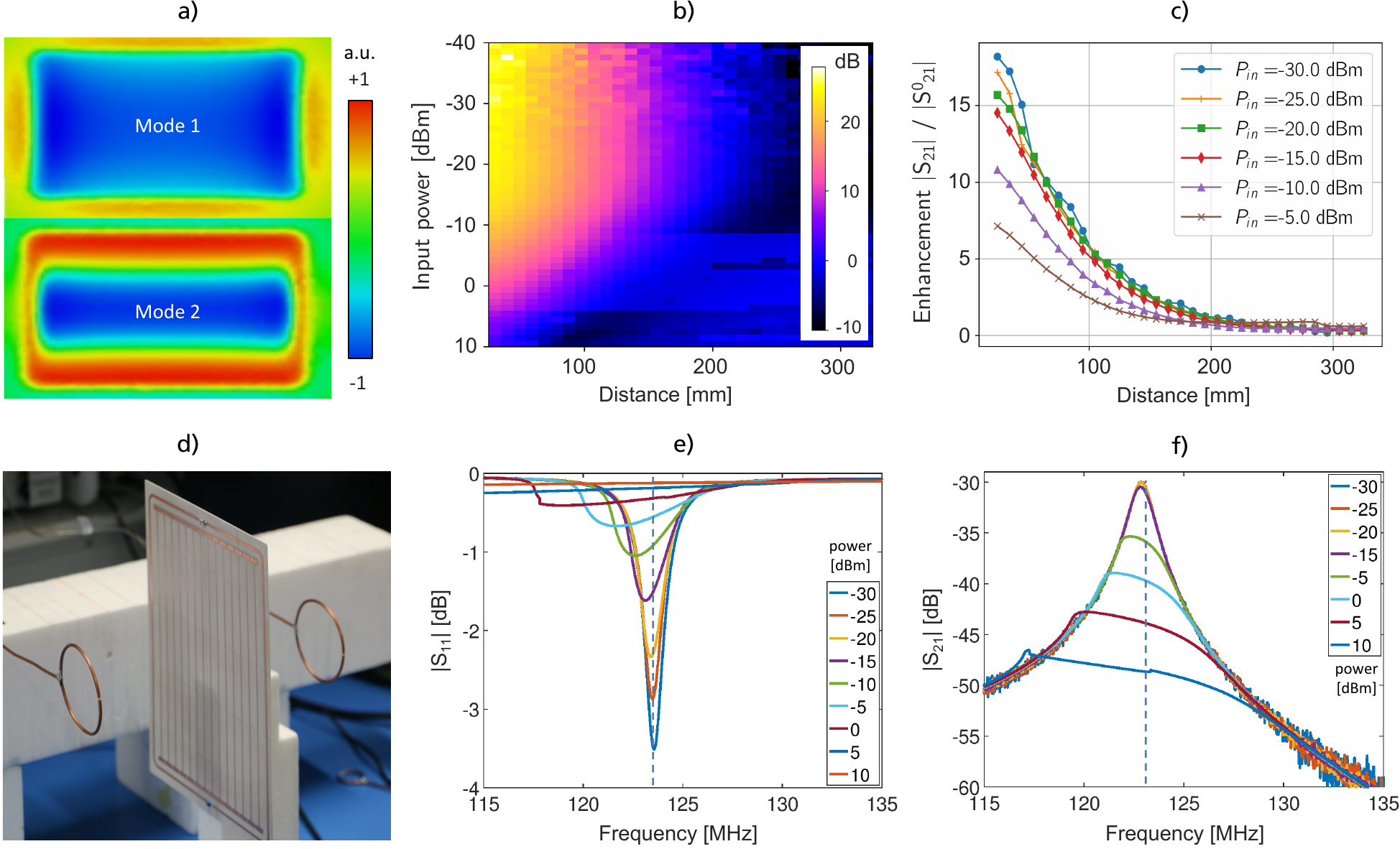}
  \caption{On-bench- characterization of the metasurfaces. a) Simulation of the first two hybrid eigenmodes (normalized H-field component orthogonal to the metasurface in arbitrary units). b) Measurement of the effective $S21$ parameter (magnitude) on the central axis as a function of distance and input power at 123.5 MHz with a 3-axis scanning stage; see the supplementary material. c) Enhancement factor curves for selected input power levels from b). d) Symmetric on-bench setup with untuned sniffer coils that leads to the scattering parameters $S11$ and $S21$ shown in e) and f) as a function of incident excitation power.}
  \label{fig:1}
\end{figure}

The characterization of the prototypes is performed on-bench with a symmetric coil setup and with a dedicated 3-axis scanning stage, respectively, see Fig.\ \ref{fig:1} and the supplementary material. 
Fine-tuning is achieved via the trimmable capacitor in the tuning resonator, which has a range of $2\,$pF to $6\,$pF (Sprague-Goodman SGC3S060). 
The on-bench measurements are performed with two untuned $60\,$mm diameter sniffer coils and a vector network analyzer (N5242A, Keysight Technologies, USA), see Fig.\ \ref{fig:1} and the supplementary material. 
The scattering S-parameters are measured as functions of the incident power to verify the desired de-tuning effect for high field strengths. 
The Q-factor of the metasurfaces was determined from S21 measurements via the width at the $3\,$dB decay from the maximum.
For EP1 it is ${Q = 110}$, and for EP2 we have ${Q = 103}$. 
Hence, EP1 performs slightly better in on-bench performance tests. However, the detuning for high incident power is smoother for EP2. 
Note that the UMX diodes have been specifically designed for MRI applications.
Slight differences in the performance can also be due to non-perfect fine-tuning and positioning accuracy but it has to be noted that both designs fulfill their purpose perfectly as they detune with increasing incident power.
The spatial dependence of the S-parameter distribution, as depicted in Fig.\ \ref{fig:1}b and \ref{fig:1}c, is determined with a workshop-built setup including a high-precision 3-axis scanning stage (SF600, GAMPT, Germany), see the supplementary material. 
This setup consists of a low-bandpass, custom-build, trimmable Tx coil with a diameter of $30\,$cm and a small Rx coil with a diameter of $2.5\,$cm, both of which are connected to a 2-Port vector network analyzer (E5061B ENA Series Network Analyzer, Keysight Technologies, USA). 
In this setup, the metasurface is mounted on a pylon of polystyrene at a distance of about $60\,$cm to the Tx coil. 
The effective scattering parameter $S21$, as depicted in Fig. \ref{fig:1}b, is measured on the central axis behind the metasurface and in planes parallel to the surface to characterize the enhancement factor vs.\ distance and the field homogeneity, respectively. 
The effective values (magnitudes) are obtained by subtracting (on a log scale) the background signal, i.e., the data from measurements without any metamaterial. 
The deduced enhancement factor is on the order of magnitude of the enhancement seen in the MRI scans.

\begin{figure}
  \centering
  \includegraphics[width=0.9\linewidth]{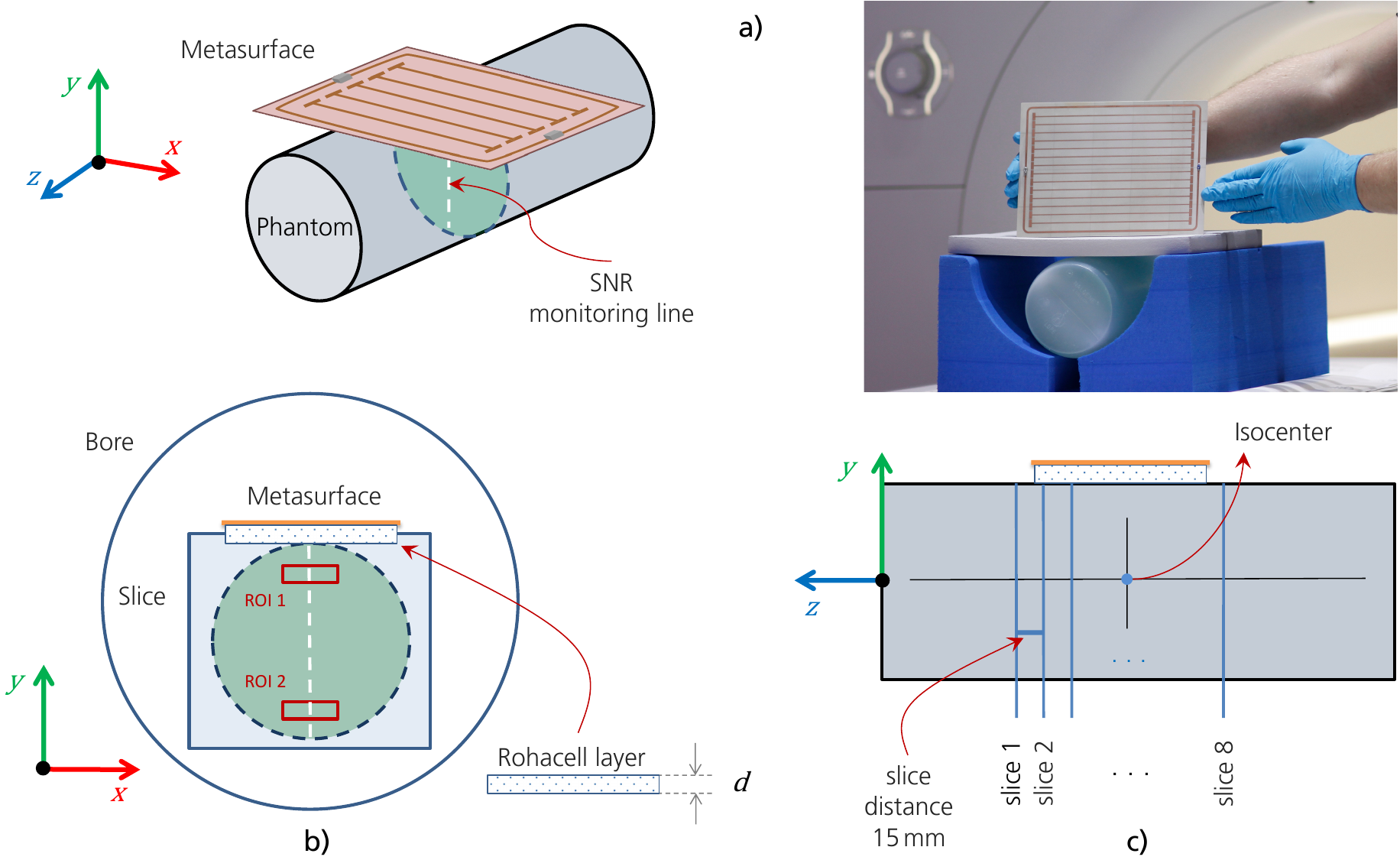}
  \caption{Orientation of the metasurface in MRI scans. a) 3D view of the setup, the photo shows the arrangement in the MRI scanning room on the patient table. b) Setup as seen from an axial plane. The SNR monitoring line and the two regions of interest are indicated. Between the metasurface and the phantom, a Rohacell layer of thickness d = 1.5 cm is used. c) Sagittal view of the orientation and recorded slices. The phantom was positioned in the isocenter. Note that for structural images with the Kiwi fruit, the slices are parallel to the enhancement plate.}
  \label{fig:2}
\end{figure}

\begin{figure}
  \centering
  \includegraphics[width=\linewidth]{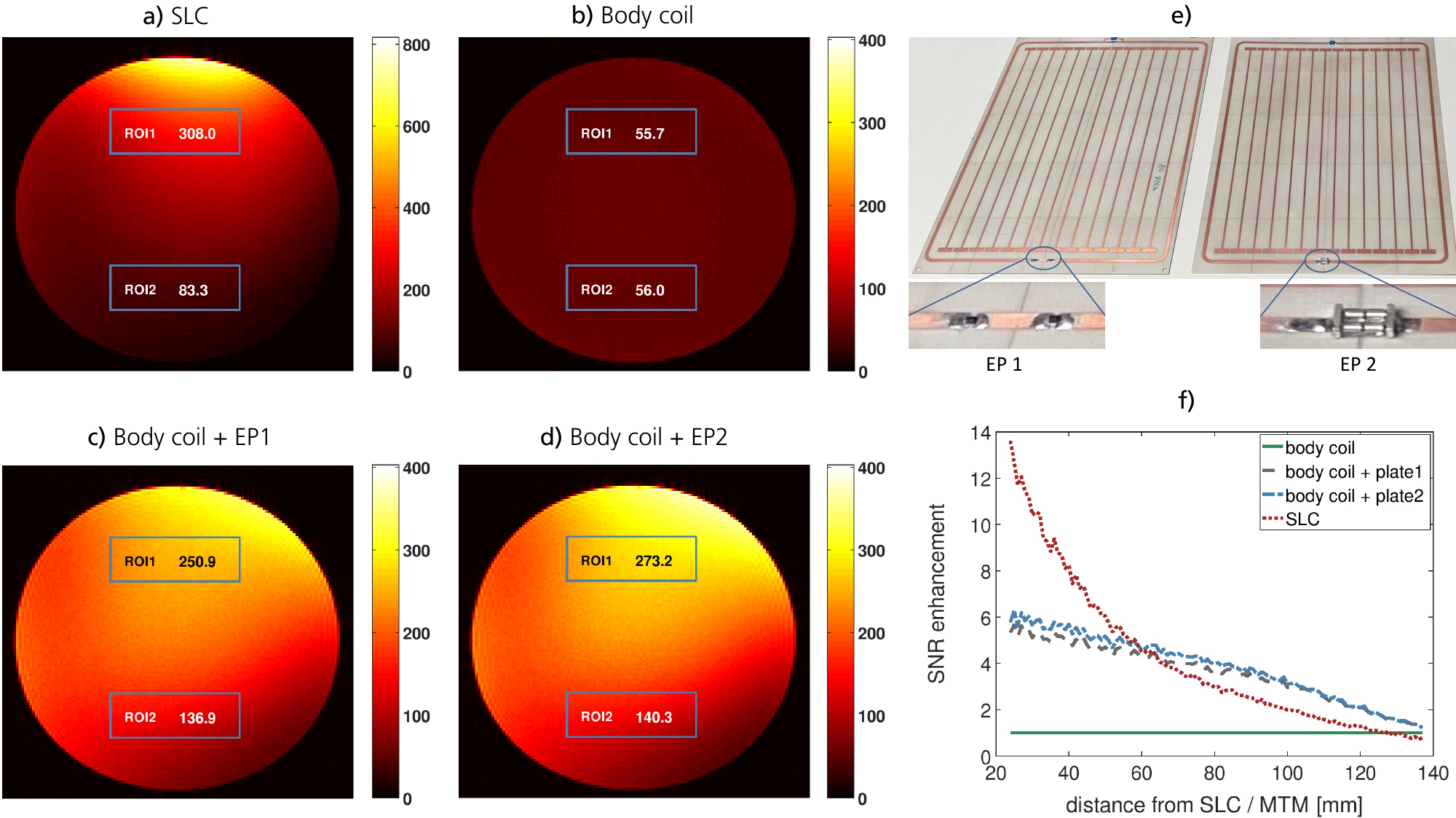}
  \caption{MRI results for homogeneous phantom scans. The SNR in slice\# 4 is shown for a flip angle of $70\,$deg (approx. Ernst angle) with Rx by a) the SLC, b) the body coil, c) body coil + EP1, and d) body coil + EP2. The SNR in the two ROIs is indicated, respectively. e) Photo of the two EPs. The insets show the elements that are responsible for detuning in Tx. f) The normalized SNR enhancement (w.r.t. the body coil) in slice\# 4 on a central vertical line, see Fig.\ \ref{fig:2}. At a certain distance, the EPs outperform the SLC. Either EP leads to a sixfold increase in SNR close to the surface, which falls off almost linearly in contrast to the SLC’s exponential behavior.}
  \label{fig:3}
\end{figure}

\begin{figure}
  \centering
  \includegraphics[width=\linewidth]{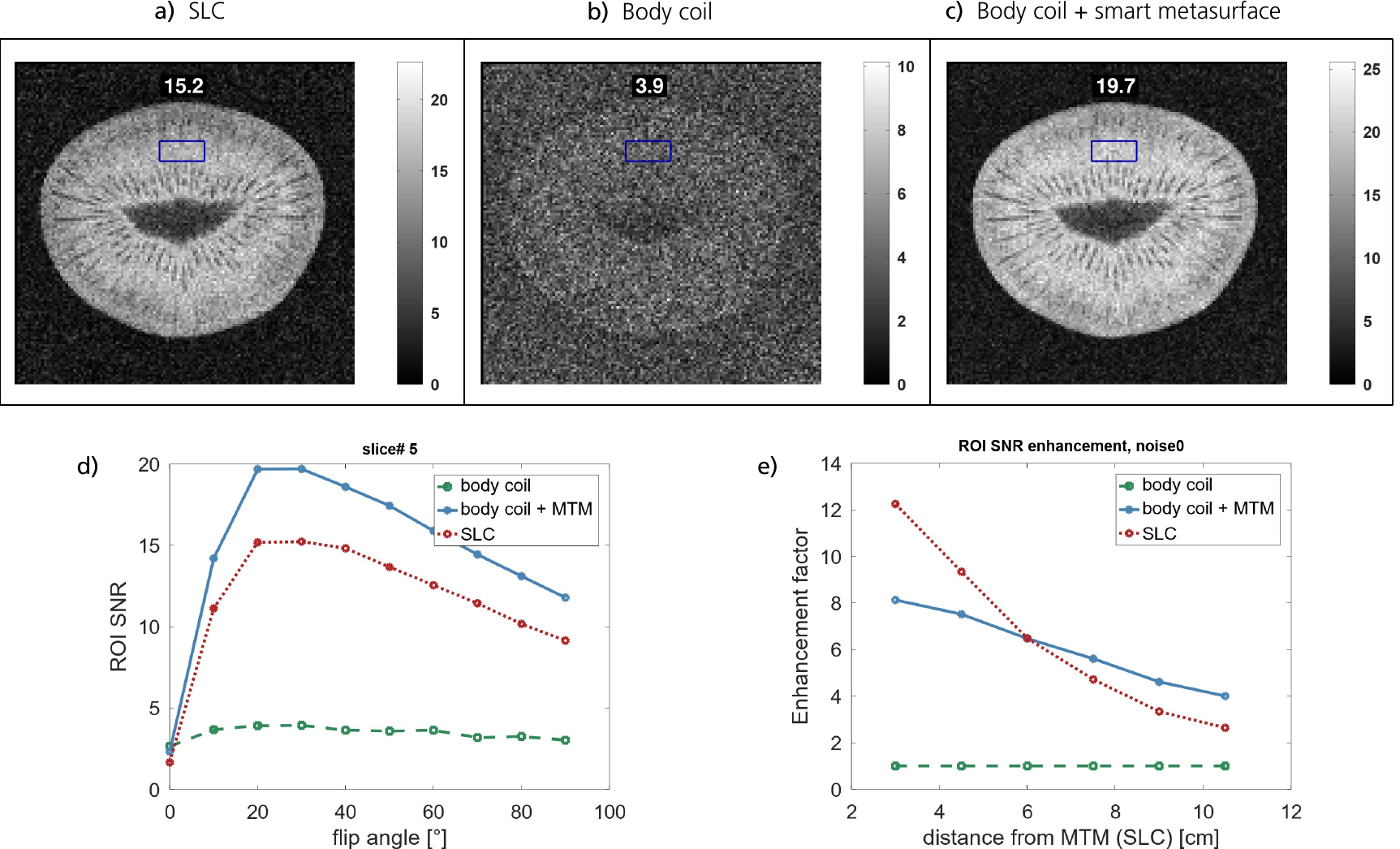}
  \caption{MRI results for structural images with EP1 using a kiwi fruit and ${TR = 100\,}$ms. The SNR is shown for slice\# 5 and Rx by a) the SLC, b) the body coil, and c) body coil + smart metasurface (EP1). Note that the slices are parallel to the metasurface. The SNR in the ROI is indicated, respectively. d) SNR in the ROI as a function of the flip angle in slice\# 5. The Ernst angle is not shifted in presence of the metasurface. Hence, the SNR increase is only due to influence on the Rx field. e) The SNR as a function of distance from the SLC / EP for the measurement with ${TR = 1\,}$s and flip angle of $75\,$deg (approx. Ernst angle). At a certain distance, the metasurface outperforms the SLC. Compared to the body coil, the metasurface leads to an eightfold increase in the SNR close to the surface.}
  \label{fig:4}
\end{figure}

\begin{figure}
  \centering
  \includegraphics[width=\linewidth]{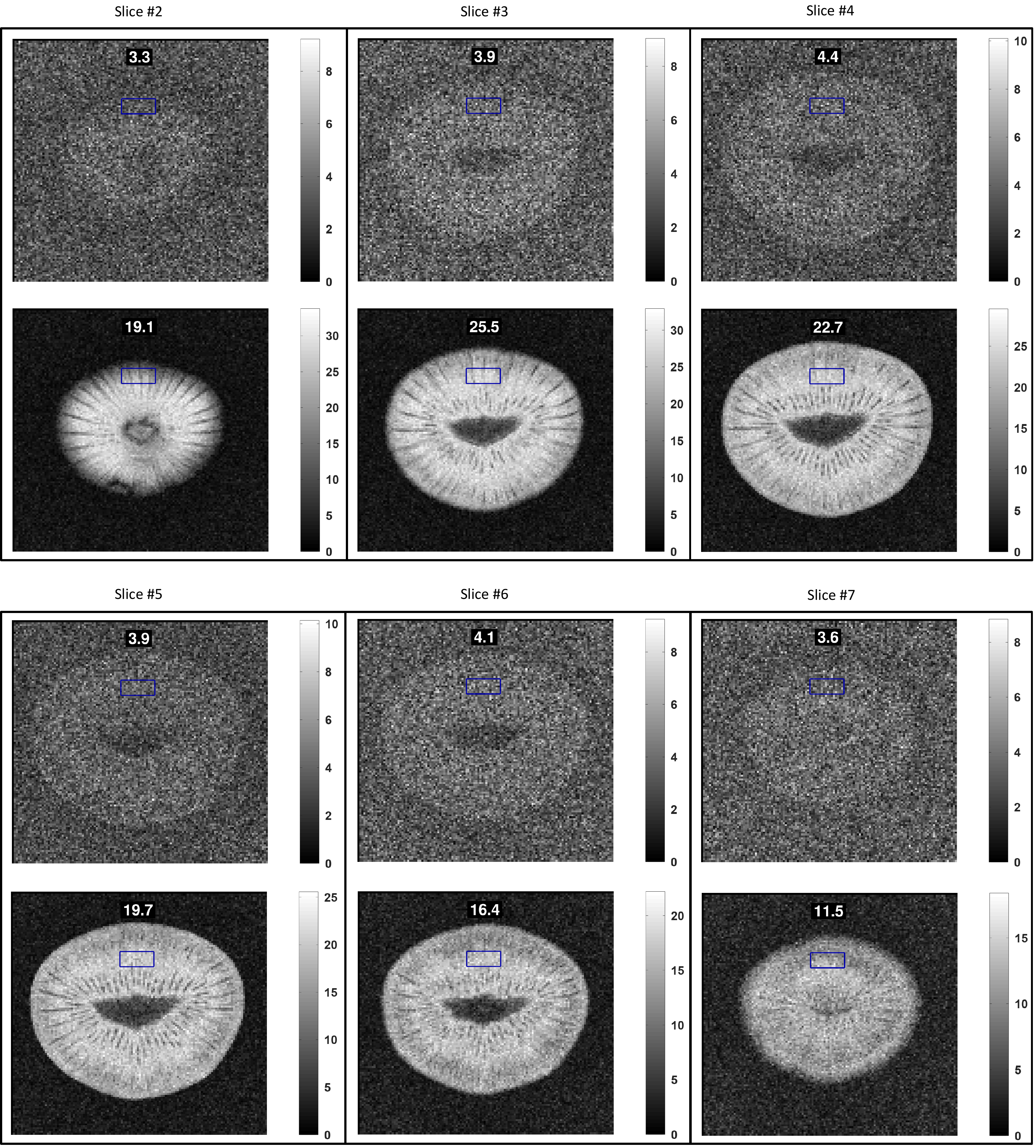}
  \caption{Additional MRI results for structural images with the Kiwi fruit and 
${TR = 100\,}$ms. The plots show the comparison of results for the body coil (top) and the body coil + smart metasurface (bottom) in each panel. The flip angle is the approx.\ Ernst angle in this configuration. The respective SNR in the ROI is indicated in the subfigures.
}
  \label{fig:5}
\end{figure}

For the MRI scans, the scanner’s body coil is used for Tx and also for Rx in combination with the smart metasurfaces, respectively. 
For comparison, we also use a $70\,$mm SLC for Rx in absence of any EP. 
Such a local coil constitutes a Rx standard to which MRI-compatible MTMs should be compared in a first step beyond the scanner's body coil.
However, such a comparison is often missing in the existing literature.
When the SLC is used, it is placed at the position of the removed EP to allow for a fair comparison.
All experiments are performed with a single sequence that offers sufficient time to place, exchange, or remove the metasurfaces between successive acquisitions without allowing the scanner to do any new adjustments in the meantime. 
In this way the genuine comparison and characterization is possible. 
We use gradient echo sequences with the following parameters (if not stated otherwise in figure captions): $TR = \{100 ms,\, 1 s\}$, $TE = 5\,$ms, field-of-view = $\{ 64\times 64, 128 \times 128 \}\,$mm$^2$, matrix size = $128 \times 128$, bandwidth = $250\,$Hz/Px, eight slices with thickness of $5\,$mm and distance of $15\,$mm. 
These sequences are implemented into the unique, vendor-independent gammaSTAR framework (Fraunhofer MEVIS)\cite{Cordes:2019,GammaStar}, which offers exceptional flexibility in sequence design and is optimally suited for MTM characterization as shown here. 
It also allows, in principle, to repeat our analysis with MRI systems from different vendors. 
Reducing the scanner’s internal image scaling factor is important. 
Even with the huge SNR enhancement factor, due to the presence of the smart metasurface, artificial image saturation should be avoided. 
We performed measurements with a homogeneous cylindrical phantom, see Figs.\ \ref{fig:2} and \ref{fig:3}, and also with a Kiwi fruit for structural images, see Figs.\ \ref{fig:4} and \ref{fig:5}. 
The supplementary material contains additional MRI results.

To prove that the smart detuning works in the Tx phase, imaging was performed for 10 different flip angles with equidistant spacing, ranging from $0\,$deg to $90\,$deg. 
In the supplementary material we show additional results (to those in Figs.\ \ref{fig:4} and \ref{fig:5}), which relate the SNR enhancement in different ROIs for the homogenous and structural phantom measurements to the nominal flip angle. 
As can be seen in all these plots, the presence of a smart metasurface does not affect the position of the maximal SNR, i.e., the position of the flip angle that leads to a maximal SNR (Ernst angle). 
Hence, the SNR improvement due to the smart metasurface is related to effects on the Rx field only. 
If the Ernst angle for any MTM in MRI measurements was shifted (as compared to body coil or SLC measurements), the MTM influences also the Tx field, which poses a potential threat to the patient. 
Uncontrolled modifications of the Tx field by MTMs are to be avoided since also all scanner-integrated safety mechanisms are not adapted to this situation.

The definition of the noise obviously influences the SNR enhancement that the metasurfaces yield in comparison to the body coil measurements. 
To characterize its effect, we evaluated the SNR performance for five different noise definitions / regions. 
The most reliable and conceptually acceptable noise measure is to calculate the standard deviation of the ‘signal’ in a large region of the $0\,$deg flip angle measurements. 
This also proves to be the most conservative measure in our consideration. For comparison, we also calculated the noise as the standard deviation in the apparently signal-free areas in the edges of each image. The results are presented in the supplementary material.

For the structural images with the Kiwi fruit and ${TR = 1\,}$s, the SNR enhancement factor due to the smart metasurface is about eight in nearby slices, see Fig.\ \ref{fig:4} and the supplementary material. 
For the measurements with ${TR = 100\,}$ms, the enhancement factor is a bit less, see Figs.\ \ref{fig:4} and \ref{fig:5}. 
However, especially in this case the influence of the metasurface is most prominent as the fruit can barely be recognized in the plots without its presence but becomes clearly visible with internal structure with a smart metasurface being used.

Note that EP2 performs slightly better in MRI experiments, see Fig.\ \ref{fig:3}.
The performance difference can be attributed to the Tx-detuning strategy via varactor and limiter diodes, respectively, but also positioning accuracy after exchange of the metasurface introduces possible differences.
The UMX limiter diodes have been specifically designed for MRI experiments but they are more expensive though.

As can be seen in Fig.\ \ref{fig:4}, the SNR enhancement due to the metasurface tends to fall off almost linearly as compared to the faster drop-off behavior of the SLC. 
This effect is also supported by simulation data, which shows that close to the surface, the enhancement can be best fitted by a linear decay. 
However, the measurements with the 3-axis scanning stage show an almost exponential decay of the enhancement vs. distance in the normal direction to the surface, see Fig.\ \ref{fig:1}. 
This difference can be explained by i) the absence of a phantom in the lab measurements and ii) the different normalization of the enhancement. 
Whilst for the MRI scans, the enhancement is calculated w.r.t.\ the body coil SNR, which is almost constant for the homogeneous phantom scans, the lab measurements yield the enhancement normalized to an empty scan, which itself has a distance-dependent behavior related to the H-field of the Tx coil.

\bibliography{improvingMRI}

\section*{Acknowledgements}

This work was supported by the Fraunhofer MAVO project MetaRF. The authors want to thank Peter Erhard for discussions and comments and Michael J\"ager for support in the production of prototypes.

\section*{Author contributions}

E.S.\ was responsible for the design and simulation of the metasurface EPs and performed on-bench characterizations. T.B., R.U.\ and J.J.\ suggested components and advised the design process. D.B.\ assisted with simulations and on-bench measurements. S.K.\ and M.G.\ drafted and performed the MRI measurements. S.K.\ developed and implemented the sequence into the gammaSTAR framework and evaluated measurement data. R.N.W.\ performed on-bench S-parameter measurements and developed the 3-axis S-parameter measurement setup. D.P.\ assisted with simulations and metasurface design, evaluated the MRI measurements, drafted the paper, and produced the figures. M.G.\ and T.B.\ initialized the research project. All authors were involved in the discussion and interpretation of the results presented and all contributed to writing the final paper. 

\section*{Additional information}

Supporting Information including extended figures and measurement data is available in the appendix.\\

\noindent The authors declare no conflict of interest.\\

\noindent The experimental research with plants (kiwi fruit used for MRI scans) complies with relevant institutional, national, and international guidelines and legislation.
The fruit was obtained from a local grocery store.

\clearpage

\clearpage
\appendix
\section*{Appendix - Supporting information}
\vspace{20pt}
\subsection*{On theoretical modeling}
An approximate theoretical description of the smart metasurfaces can be obtained using the coupled mode theory.
For the linear part of the system, the inner metasurface structure, the mode amplitude $y_1$ depends on an excitation $x$ according to
\begin{align}
	\dot{y}_1 = a_1(\omega_1) \, y_1 + b_1 \, x \, , \quad a_1(\omega_1) = i \omega_1 - 1/\tau_r - 1/\tau_i \, , \quad b_1 = \sqrt{2/\tau_r} \, .
\end{align}
Here, $\omega_1$ is the resonance frequency, $\tau_r$ is the decay time due to radiation losses, and $\tau_i$ is the decay time associated with intrinsic losses. The overdot denotes the time derivative.
Note that the inverse of the lifetimes $\tau_i$ are to be considered as decay rates.
The coupling to the excitation is via the radiative term.
A similar equation can be constructed for the tuning resonator,
\begin{align}
	\dot{y}_2 = a_2(\omega_2) \, y_2 + b_2 \, x \, ,
\end{align}
but now the resonance frequency $\omega_2$ depends on the resonator's amplitude as well, $\omega_2 = \omega_2(y_2)$.
We may assume a linear relation such that $\omega_2 = \Omega_2 + \kappa |y_2|$ with a constant $\kappa$ that depends on the properties of the chosen non-linear element and a small amplitude limit $\Omega_2$.
Finally, the full smart metasurface is described by the coupled, non-linear system of differential equations in the time domain,
\begin{subequations}
\begin{align}
	\dot{y}_1 &= \big( i \omega_1 - 1/\tau_{r,1} - 1/\tau_{i,1} \big) y_1 + \sqrt{2/\tau_{r,1}} \, x + i k y_2 \, , \\
	\dot{y}_2 &= \big( i \Omega_2 + i\kappa |y_2| - 1/\tau_{r,2} - 1/\tau_{i,2} \big) y_2 + \sqrt{2/\tau_{r,2}} \, x + i k y_1 \, ,
\end{align}
\end{subequations}
which includes the coupling strength $k$.
Let us be precise regarding the dimension of involved quantities.
Certainly, $\omega$ is an inverse time, i.e., it is measured in Hz.
All $\tau_i$ are times, measured in seconds. 
For the excitation $x$ we have that the dimension of $|x|^2$ is a power, thus for the mode amplitude the dimension of $|y_i|^2$ is an energy.
Hence, the dimension of the quotient of mode amplitude and excitation is $\big[ |y_i|^2 / |x|^2 \big] = \,$s.  
The system of equations can be simplified by assuming that the excitation is expressed as $x = |x| \exp(i\omega t)$.
Then, by proceeding to the frequency domain the time derivatives are converted into algebraic expressions and the equations decouple such that
\begin{multline}
	\left[ \big(i (\omega - \omega_1) + 1/\tau_{r,1} + 1/\tau_{i,1} \big) \big( i(\omega - \Omega_2 - \kappa |y_2|) + 1/\tau_{r,2} + 1/\tau_{i,2} \big) + k^2 \right] y_2 \\
	  = \left[ i k \sqrt{2/\tau_{r,1}} + \left( i(\omega-\omega_1) + 1/\tau_{r,1} + 1/\tau_{i,1} \right) \sqrt{2/\tau_{r,2}} \right] |x| 
\end{multline}
remains to be solved numerically for $y_2$, and $y_1$ is subsequently given by
\begin{align}
	y_1 = \left( i k y_2 + \sqrt{2/\tau_{r,1}} \, |x| \right) \big( i(\omega - \omega_1) + 1/\tau_{r,1} + 1/\tau_{i,1} \big)^{-1} \, .
\end{align}
To calculate the reflection of the full system, we use
\begin{align}
	R = \dfrac{\left|-x + \sqrt{2/\tau_{r,1}} \, y_1 + \sqrt{2/\tau_{r,2}} \, y_2 \right|^2}{|x|^2} \, .
\end{align}
This equation can be used to model our measurements of the S11 parameter with the untuned sniffer coil.
The theoretical description fits the experimental results qualitatively when the eight parameters are chosen appropriately.
For small excitation strength, there is only one solution for the mode amplitude $y_2$, whereas for strong excitations three solution branches appear, of which the middle one is unstable, see Fig.\ S\ref{figExt:1}.
Note that all parameters needed for the model, i.e., resonance frequencies, decay rates, and couplings can be given as multiples of the fundamental resonance $\omega_1$.
These eight parameters are the fundamental resonances of the two coupled systems $\omega_1,\omega_2$, the coupling strength $k$ between the two, a parameter $\kappa$ to describe the non-linear behaviour of the tuning resonator, and the time scales $(t_{r,1},t_{r,2}), (t_{i,1},t_{i,2})$ related to radiation losses and internal losses, respectively.
They have to following values for the model of the reflection $R$, which is shown in supplementary Fig.\ S\ref{figExt:1}b.
\begin{subequations}
\begin{align}
	t_{r,1} 	&= 3000 / \omega
	_1 \, , \quad t_{r,2} = 100 / \omega_1 \\
	t_{i,1} 	&=  600 / \omega_1\, , \quad t_{i,1} = 100 / \omega_1\\
	\omega_1 	&= 136 \,\, \text{MHz} \, , \quad \Omega_2 = \omega_1 + 2 \,\, \text{MHz}\\
	\kappa 		&= -0.16 \omega_1 \, , \quad k = 0.1  \omega_1 \, .
\end{align}
\end{subequations}
To model the magnetic field focussing effect, i.e., the SNR enhancement, the following steps are needed in principle.
(i) The current distribution in individual wire-resonator unit cells and the tuning resonator is to be modeled / numerically solved for.
(ii) Then, using Bio-Savart's law the induced magnetic field at some distance can be calculated.
(iii) The obtained magnetic field is to be superimposed on the excitation field and the enhancement is given by
\begin{align}
	\eta(\vec{x}) := \dfrac{|B_{\text{ind},z}(\vec{x};\omega) + B_{0,z}(\vec{x})|}{|B_{0,z}(\vec{x})|} \, ,
\end{align}
in which $B_{\text{ind},z}(\vec{x};\omega)$ is $z-$component of the induced magnetic field, which is extremal at the design resonance frequency.
In future work, a detailed numerical and mathematical study of this and other (advanced) metasurfaces will be presented including quantitative matching of theoretically modeled, simulated, and measured system characteristics as well as simulations of the signal focusing in MRI experiments.

\clearpage
\subsection*{Additional Figures and Data}
\setcounter{figure}{0}   
\renewcommand\figurename{Figure S}

\begin{figure}[h!]
  \centering
  \includegraphics[width=\linewidth]{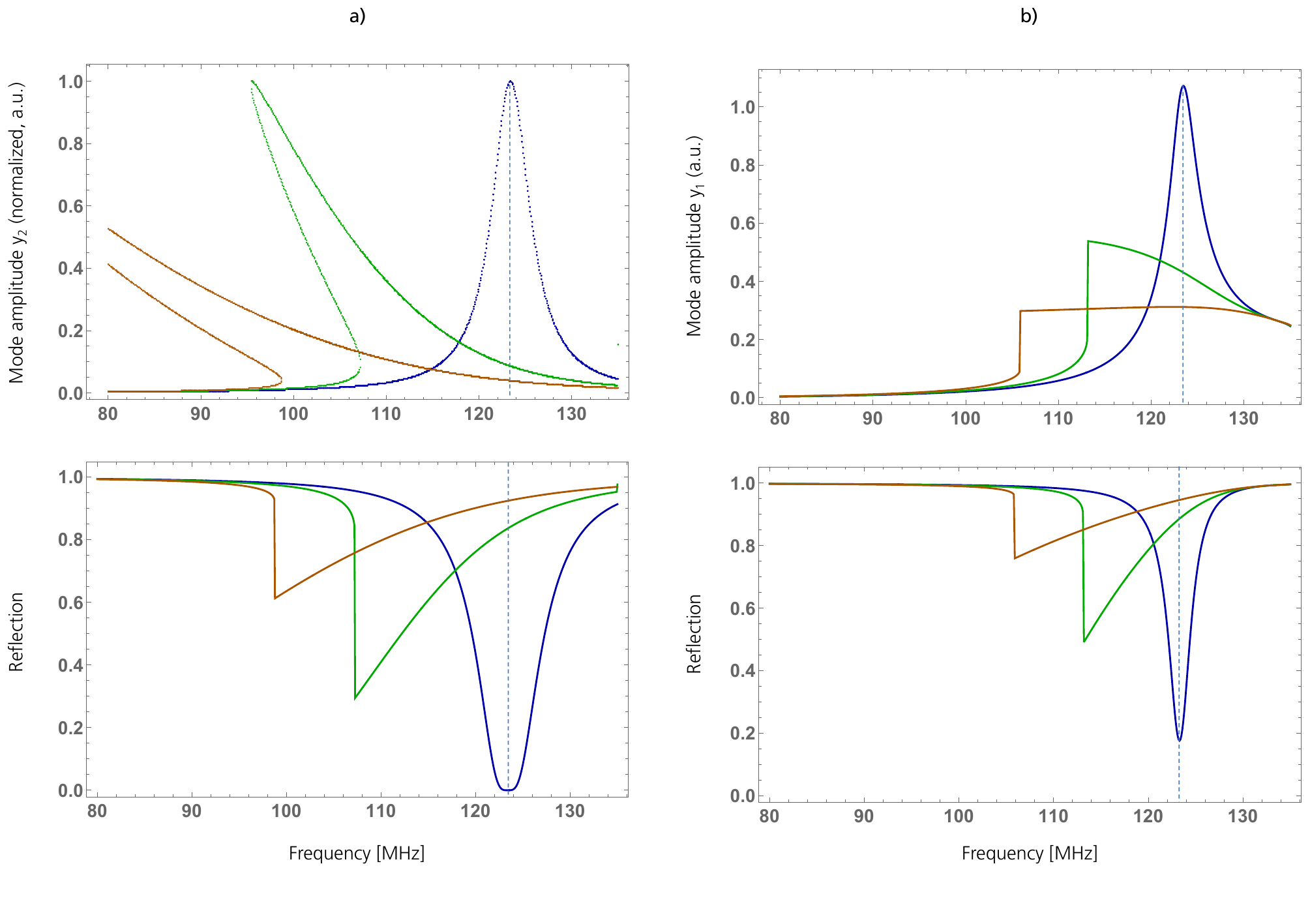}
  \caption{Theoretical modeling of the non-linear behaviour for different excitation strengths. a) The mode amplitude $y_2$ and the reflection $R$ of the non-linear resonator are shown for three different incident power levels (isolated tuning resonator with $\Omega_2 = 123.5\,$MHz, without coupling, i.e., $k=0$). b) Mode amplitude $y_1$ and reflection $R$ of the coupled, non-linear metasurface system; compare Fig.\ 1e in the main text, which shows the S11 measurements with an untuned sniffer coil. For low incident power (blue curves), the structure is resonant at the MR scanner's resonance frequency. At higher incident power (green, brown) the system becomes progressively detuned.
}
  \label{figExt:1}
\end{figure}

\begin{figure}
  \centering
  \includegraphics[width=\linewidth]{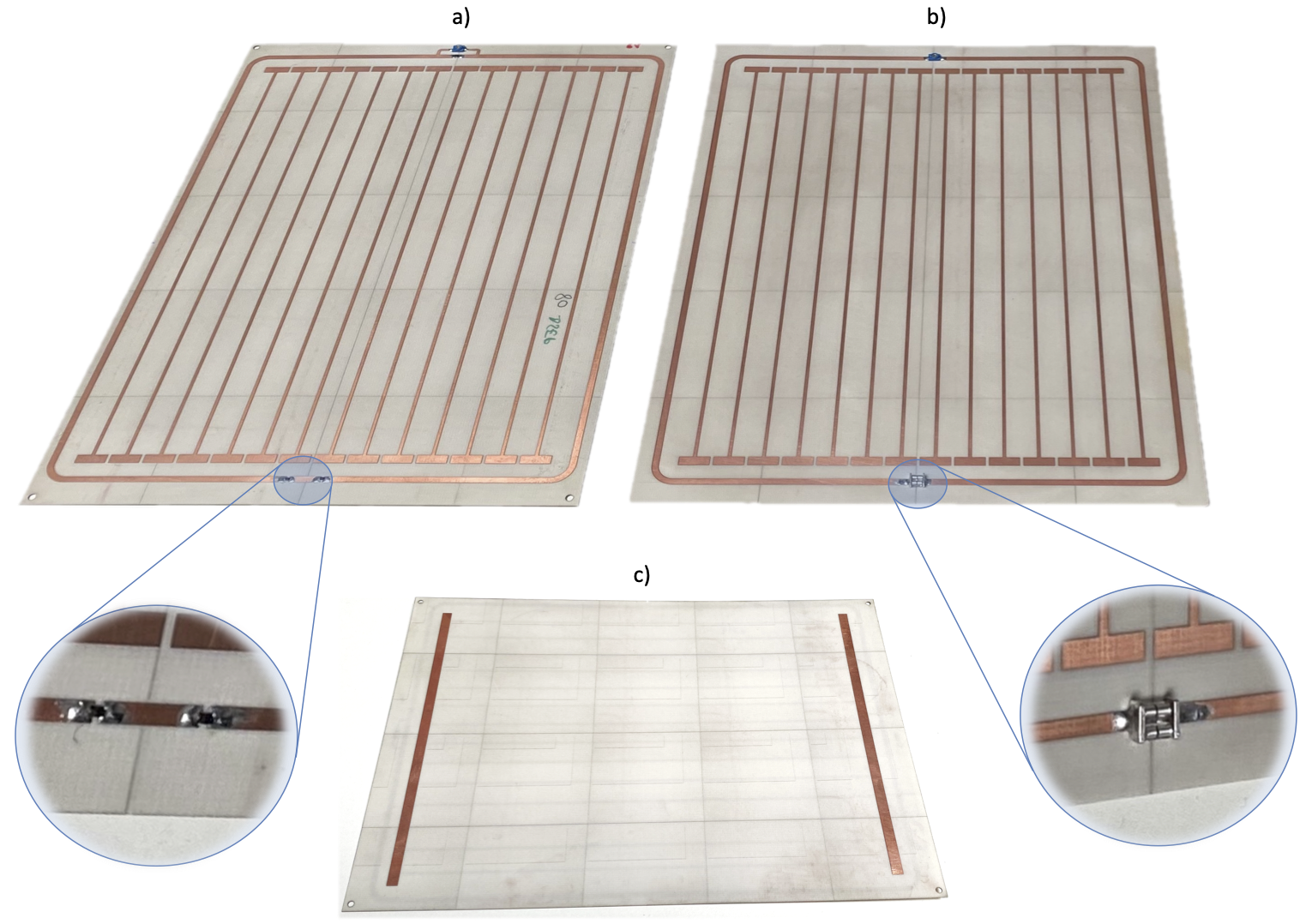}
  \caption{The two manufactured smart metasurface enhancement plates with a) varactor-loaded tuning resonator and b) limiter-diode-loaded tuning resonator. c) The back of the plate with the ground patches for capacitively coupled wire resonator unit cells is shown, which is the same for both EPs. The ground stripes on the back in combination with the rectangular patches at the end of the wire-resonators on the front form parallel plate capacitors and couple the individual unit cells.
}
  \label{figExt:2}
\end{figure}

\begin{figure}
  \centering
  \includegraphics[width=\linewidth]{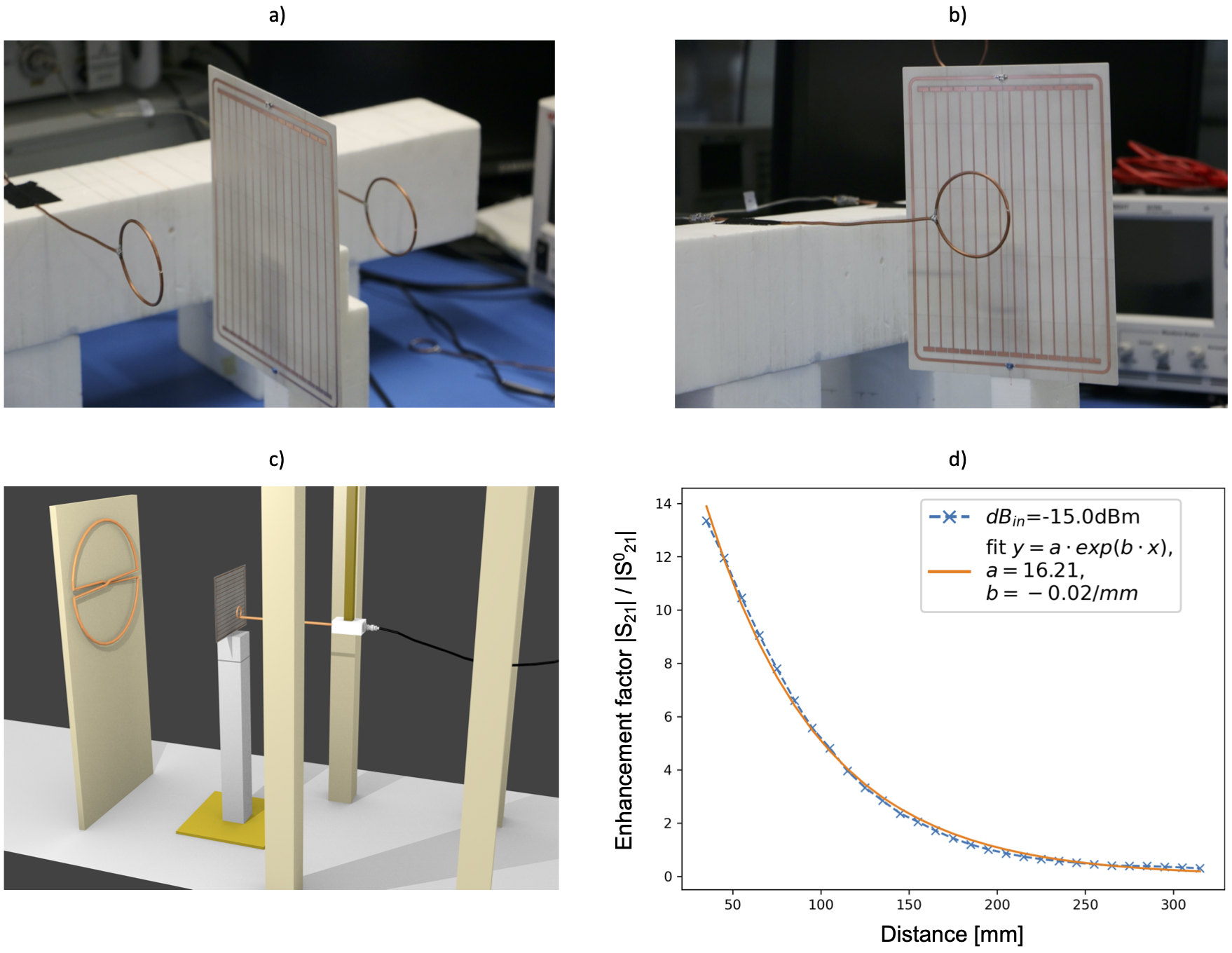}
  \caption{On-bench characterization of the manufactured prototypes using un-tuned sniffer coils and a vector network analyzer. a) Two sniffer coils positioned for $S21$ measurements. b) A single sniffer coil closer to the EP for $S11$ measurements. c) Schematic of the setup used for measurements of the spatial dependence of scattering parameters. d) Fit of the exponential decay of the enhancement factor as a function of the distance. At $-15\,$dBm input power, we observe an exponential decay of about 0.02/mm.
}
  \label{figExt:3}
\end{figure}

\begin{figure}
  \centering
  \includegraphics[width=\linewidth]{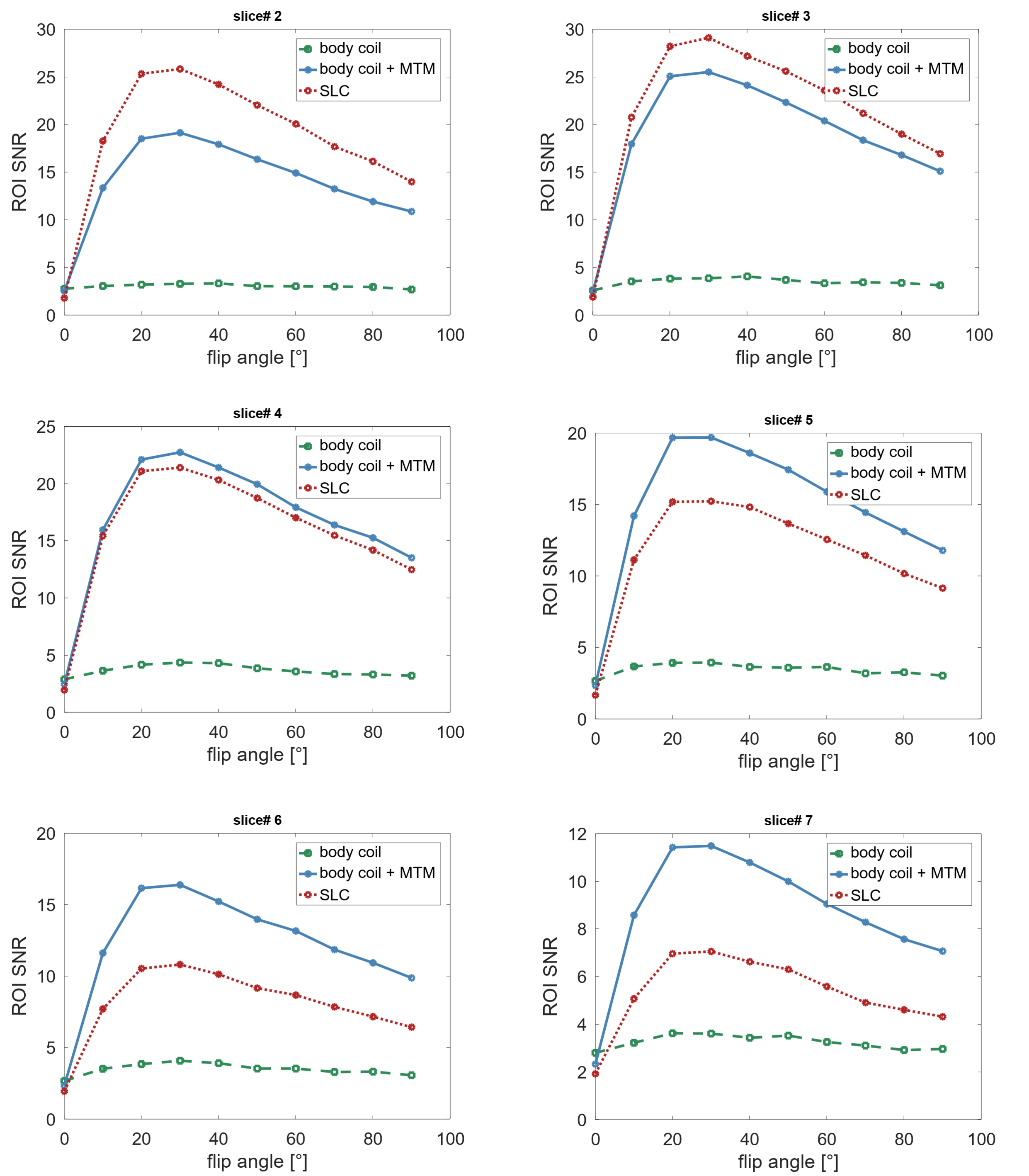}
  \caption{Additional MRI results for structural images with the Kiwi fruit with ${TR = 100\,}$ms. The plots show the SNR in the ROI in different slices (see Fig. 4). The slices are parallel to the metasurface, i.e., the slice number is a measure of distance. The maximum’s position is unchanged by the presence of the metasurface in all slices, respectively. This indicates that the EP only influences the Rx field.
}
  \label{figExt:4}
\end{figure}

\begin{figure}
  \centering
  \includegraphics[width=\linewidth]{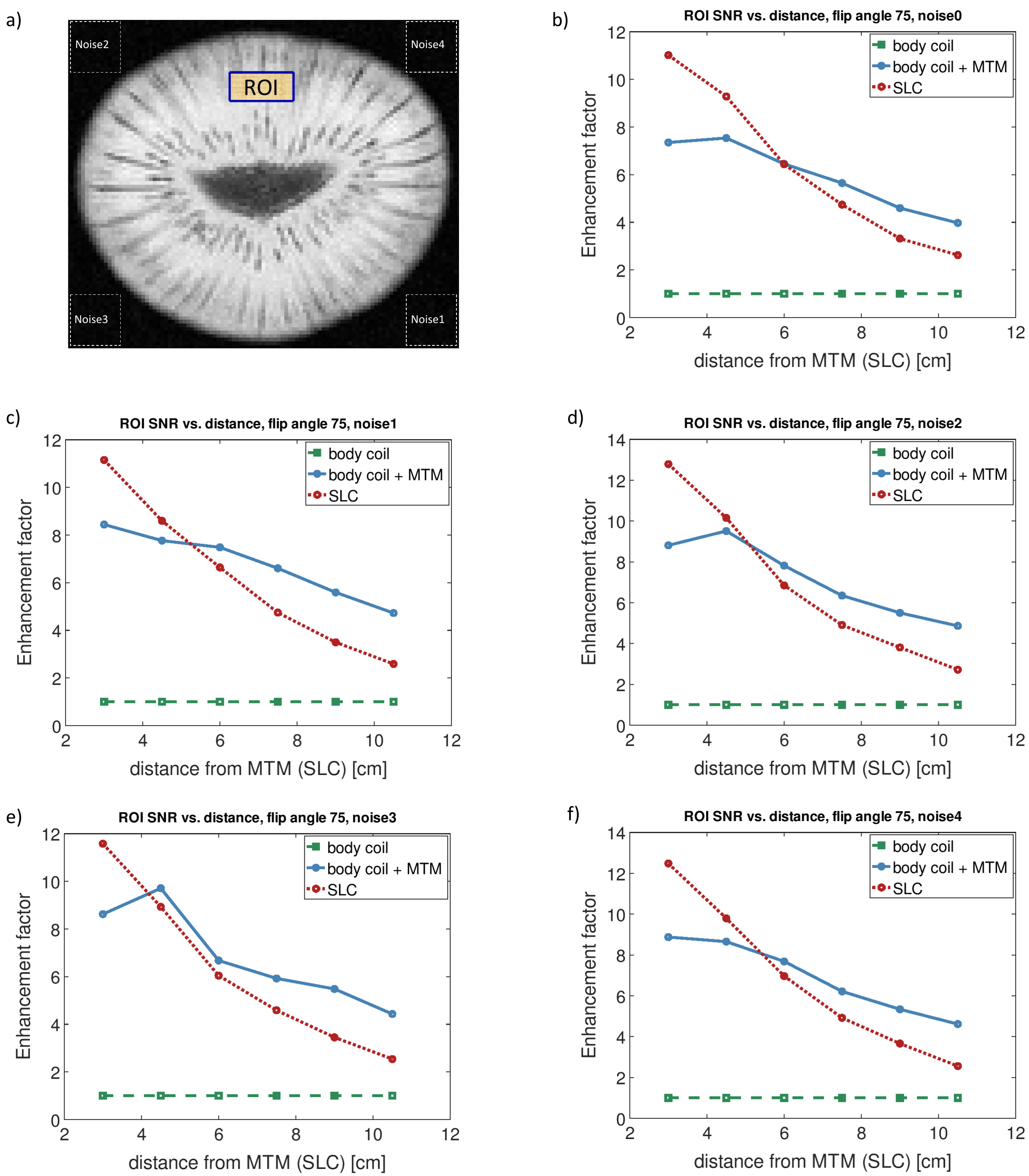}
  \caption{Additional MRI results for structural images with the Kiwi fruit with ${TR = 1\,}$s. The plots show the normalized (w.r.t.\ the body coil) SNR in the ROI vs.\ distance from the metasurface for different noise definitions. For noise0, the noise is calculated from the $0\,$deg flip angle measurements, see the methods section. In all other cases, the noise is the standard deviation in the indicated apparently ‘signal-free’ areas, respectively. The smart metasurface enhancement factor drops off slower with increasing distance as compared to the SLC. As can be seen, it depends, of course, on the definition of ``noise'' but the noise as calculated from the $0\,$deg flip angle scans gives the most conservative measure.
}
  \label{figExt:5}
\end{figure}

\begin{figure}
  \centering
  \includegraphics[width=0.87\linewidth]{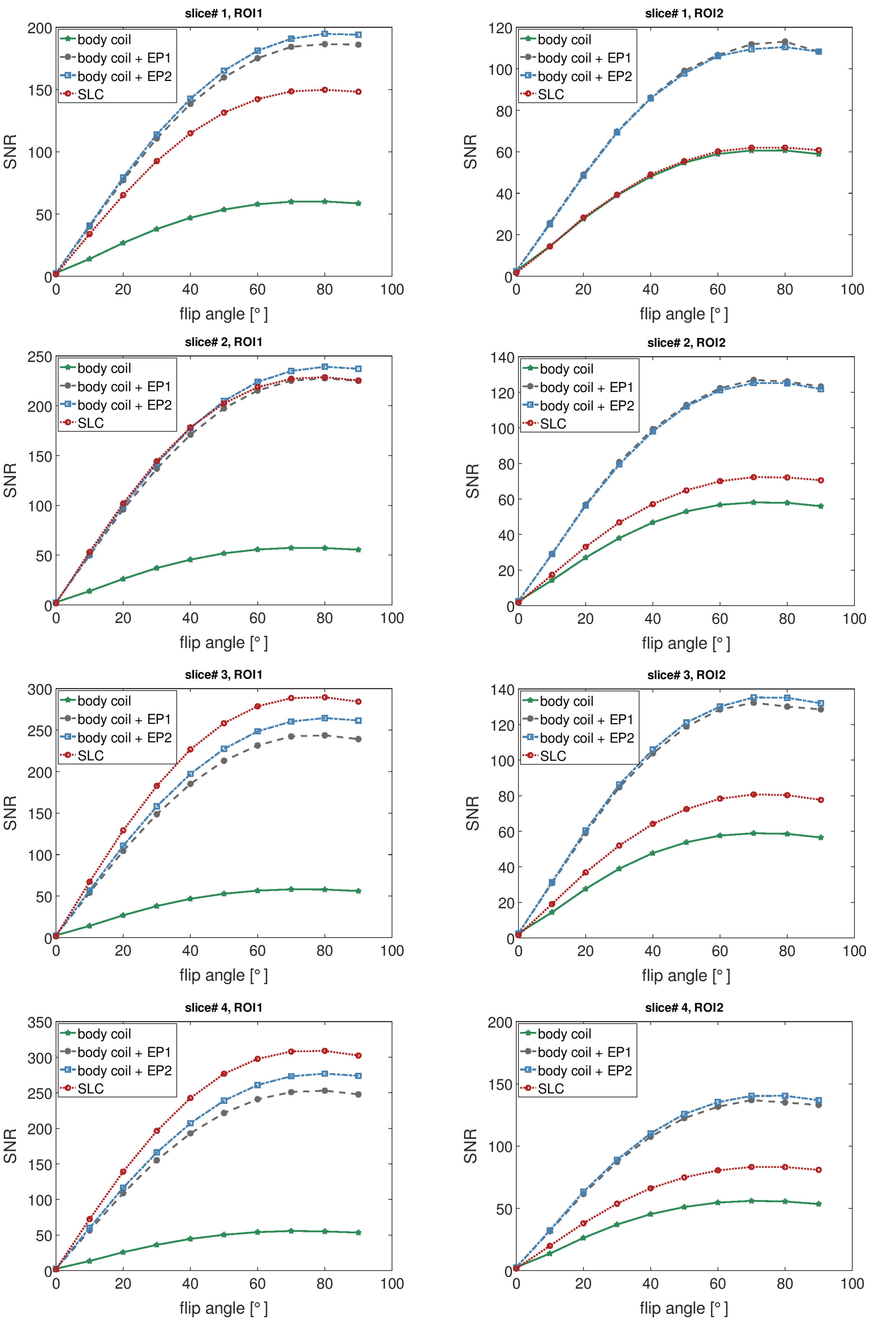}
  \caption{Additional MRI data for phantom measurements. The SNR is shown in the two ROIs as a function of the flip angle for slices\# 1--4. The maximum’s position (Ernst angle) is unchanged in presence of either EP, thus, the smart metasurfaces do not influence the Tx field and the SNR increase is purely due to Rx effects.
}
  \label{figExt:6}
\end{figure}

\begin{figure}
  \centering
  \includegraphics[width=0.87\linewidth]{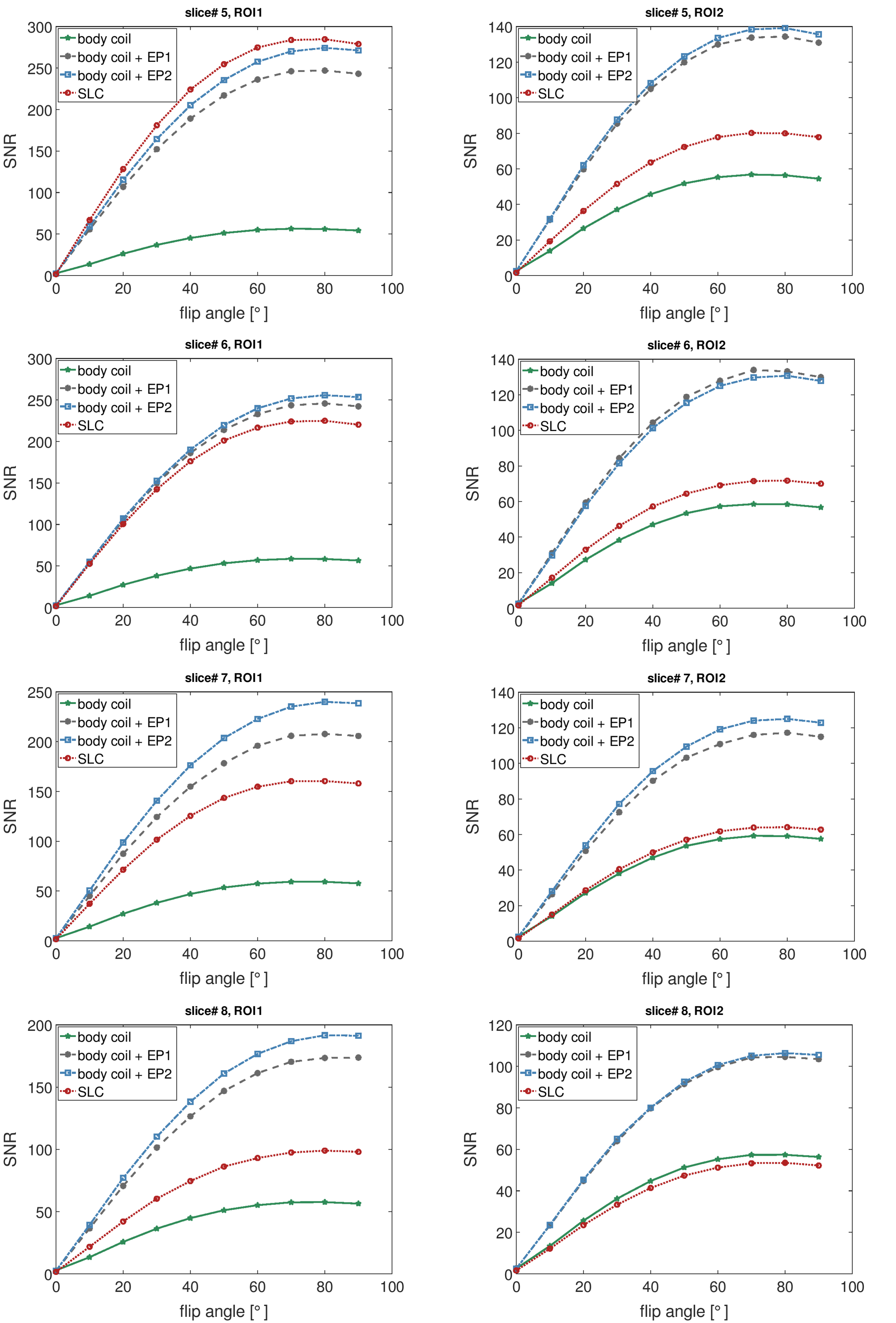}
  \caption{Extension of the previous figure for slices\# 5--8.
}
  \label{figExt:7}
\end{figure}

\end{document}